\documentclass[aps, prl, twocolumn, lengthcheck, superscriptaddress, 
nofootinbib]{revtex4-1}

\usepackage{amsmath}
\usepackage{graphicx}
\usepackage{color}
\usepackage{times}
\usepackage{inputenc}
\usepackage{bm}
\usepackage{multirow}
\usepackage{float}
\usepackage{url}
\usepackage{natbib}
\usepackage{tensor}
\usepackage[colorlinks=true,citecolor=blue,urlcolor=blue,linkcolor=red]{hyperref}
\usepackage[normalem]{ulem}
\usepackage{subfigure}

\def\fm3{\;\text{fm}^{-3}}

\begin{document}
\title{
Implication of neutron star observations to the origin of nucleon mass}

\author{Bikai Gao}
\email{bikai.gao@gmail.com}
\affiliation{Research Center for Nuclear Physics (RCNP), Osaka University, Osaka 567-0047, Japan}

\author{Xiang Liu}
\affiliation{Department of Physics, Nagoya University, Nagoya 464-8602, Japan}

\author{Masayasu Harada}
\email{harada.masayasu.j7@f.mail.nagoya-u.ac.jp}
\affiliation{Kobayashi-Maskawa Institute for the Origin of Particles and the Universe, Nagoya University, Nagoya, 464-8602, Japan}
\affiliation{Department of Physics, Nagoya University, Nagoya 464-8602, Japan}
\affiliation{Advanced Science Research Center, Japan Atomic Energy Agency, Tokai 319-1195, Japan}

\author{Yong-Liang Ma}
\email{ylma@nju.edu.cn}
\affiliation{School of Frontier Sciences, Nanjing University, Suzhou, 215163, China}

\date{\today}

\begin{abstract}
We investigate the implications of neutron star observations for understanding the origin of nucleon mass using a framework that combines three complementary approaches: the equation of state based on parity doublet structure for hadronic matter below $2n_0$, the Nambu-Jona-Lasinio (NJL) model for quark matter above $5n_0$, and a model-independent analysis of the intermediate density region based on fundamental physical principles. By systematically exploring parameter spaces and comparing theoretical predictions with recent observational constraints, we establish constraints on the chiral invariant mass. Our results suggest that more than a half of the nucleon mass originates from sources beyond spontaneous chiral symmetry breaking, challenging conventional understanding of nucleon mass generation.  These constraints arise solely from fundamental physical principles and observational data, independent of specific assumptions about the nature of the quark-hadron transition, providing robust insights into the microscopic origin of hadron masses.
\end{abstract}

\maketitle 

\textit{Introduction.---}
Protons and neutrons (collectively called nucleons) are composed of quarks and gluons, which are described by Quantum Chromodynamics (QCD)---the fundamental theory of strong interactions. Nucleon has the typical mass about 1000 MeV while its components---three current quarks---only possess mass about 10 MeV. More than 98$\%$ of nucleon mass is generated by QCD dynamics, and understanding this mass generation mechanism is one of the most important subjects in hadron physics.

One candidate for the origin of nucleon (hadron) mass is quark-antiquark condensate $\langle\bar{q}q\rangle$, which triggers spontaneous breaking of chiral symmetry and generates mass dynamically. The second is the chiral invariant mass, whose existence also supported by the lattice QCD calculations \cite{Aarts:2015mma,Aarts:2017rrl,Aarts:2018glk} and QCD sum rule calculations~\cite{Kim:2020zae,Kim:2021xyp,Lee:2023ofg}.
This component remains independent of chiral symmetry and originates mainly from gluon dynamics. While the former has been extensively studied in various theoretical frameworks \cite{Gasser:1983yg,Hatsuda:1994pi,Hayano:2008vn,Bazavov:2011nk,Fukushima:2013rx,Buballa:2014tba,Gubler:2018ctz}, the latter component has been frequently overlooked. One approach to explicitly incorporate both the chiral invariant mass and the component arising from spontaneous chiral symmetry breaking in the study of hadron mass generation is to use the parity doublet structure~\cite{Detar:1988kn, Jido:2001nt}. In this framework, excited nucleons with negative parity, such as $N^*(1535)$, are considered as chiral partners of the ground-state nucleons. Currently, it remains unknown which of these components dominates, what proportion each contributes to nucleon mass, or whether additional origins exist.

Inside highly dense objects such as neutron stars (NS), chiral symmetry gradually restores, and the nucleon mass from the quark condensation disappear, remaining only the chiral invariant mass. This makes NS a natural laboratory to study the origin of nucleon mass. Precise measurements from multiple observational channels—including pulsar timing, gravitational wave detections from binary NS mergers, and X-ray observations of NS radii \cite{Fonseca:2016tux,LIGOScientific:2017vwq,LIGOScientific:2017ync,Miller:2021qha,Riley:2021pdl,Fonseca:2021wxt,Vinciguerra:2023qxq,Kacanja:2024hme}—have placed increasingly stringent constraints on the NS equation of state (EOS). This EOS should directly related to the chiral invariant mass  $m_0$ and chiral dynamics, which governs the stiffness of nuclear matter at densities relevant to NS cores. By systematically comparing theoretical mass-radius ($M$-$R$) relationships derived from different values of the chiral invariant mass against observational constraints, we can establish new bounds on the chiral invariant mass and provide important clue to understand the hadron mass generation mechanism.

To construct the NS EOS, a key challenge lies in describing matter across different density regimes. At relatively low densities ($n_B \lesssim 2n_0$ with $n_0 \approx 0.16$~fm$^{-3}$ being normal nuclear density), matter exists in the hadronic phase where chiral effective models such are applicable. At higher densities ($n_B \gtrsim 5n_0$), quarks may become deconfined, requiring consideration of quark degrees of freedom. However, the intermediate density region ($2n_0 \sim 5n_0$) the description of NM is particularly uncertain. As nuclear interactions become increasingly complex while quark degrees of freedom may gradually emerge \cite{McLerran:2007qj,McLerran:2008ua,Kojo:2021ugu,Fujimoto:2023mzy,Gao:2024jlp}. Traditional approaches bridge this intermediate gap through extrapolation or interpolation methods, often assuming specific scenarios of the quark-hadron transition, such as a smooth crossover \cite{Masuda:2012ed,Masuda:2012kf, Baym:2017whm, Baym:2019iky,Kojo:2021wax, Minamikawa:2020jfj,Blaschke:2021poc,Minamikawa:2023eky,Kong:2025dwl} or a first-order phase transition \cite{Lenzi:2012xz,Benic:2014jia,Contrera:2022tqh,Christian:2023hez,Gao:2024lzu,Li:2024sft,Yuan:2025dft}. It should be noted that, quark is not an indispensable degree of freedom in this density region. Some models with only unflavored hadrons can also describe NM properties saturate the constraints from NS observations \cite{Ma:2018xjw,Ma:2019ery,Li:2022okx,Miyatsu:2022wuy,Zhang:2024sju,Zhang:2024iye,Gholami:2024ety}. A common problem shared in modeling NM in the intermediate density region is that there are more or less model dependencies difficult to quantify.

In this work, we develop a novel approach that minimizes model dependencies in the intermediate density region by combining three complementary descriptions: the EOS based on parity doublet structure for the low-density hadronic phase \cite{Zschiesche:2006zj,Dexheimer:2007tn,PhysRevC.77.025803,PhysRevC.82.035204,Steinheimer:2011ea,Dexheimer:2012eu,Motohiro:2015taa,Marczenko:2020jma,Marczenko:2022hyt,Marczenko:2024jzn,Marczenko:2024nge,Yasui:2024dbx} up to $2n_0$, the NJL model for the high-density quark matter phase above $5n_0$ \cite{Hatsuda:1994pi,Baym:2017whm, Baym:2019iky, Kojo:2021wax, Yuan:2023dxl, Gao:2024chh, Gao:2024lzu,Gholami:2024ety}, and a model-independent analysis of the intermediate region based solely on fundamental physical principles including causality and thermodynamic stability as discussed in \cite{Komoltsev:2021jzg,Somasundaram:2022ztm,Komoltsev:2023zor,Kurkela:2024xfh}.

By systematically exploring the parameter spaces, we examine the implications of recent NS observations for understanding the origin of nucleon mass, bridging microscopic nuclear physics with macroscopic NS properties. Our investigation leads to constraints that arise solely from fundamental physical principles, independent of specific assumptions about the nature of the hadron-quark transition.

\begin{figure}[htbp]\centering
\includegraphics[width=1\hsize]{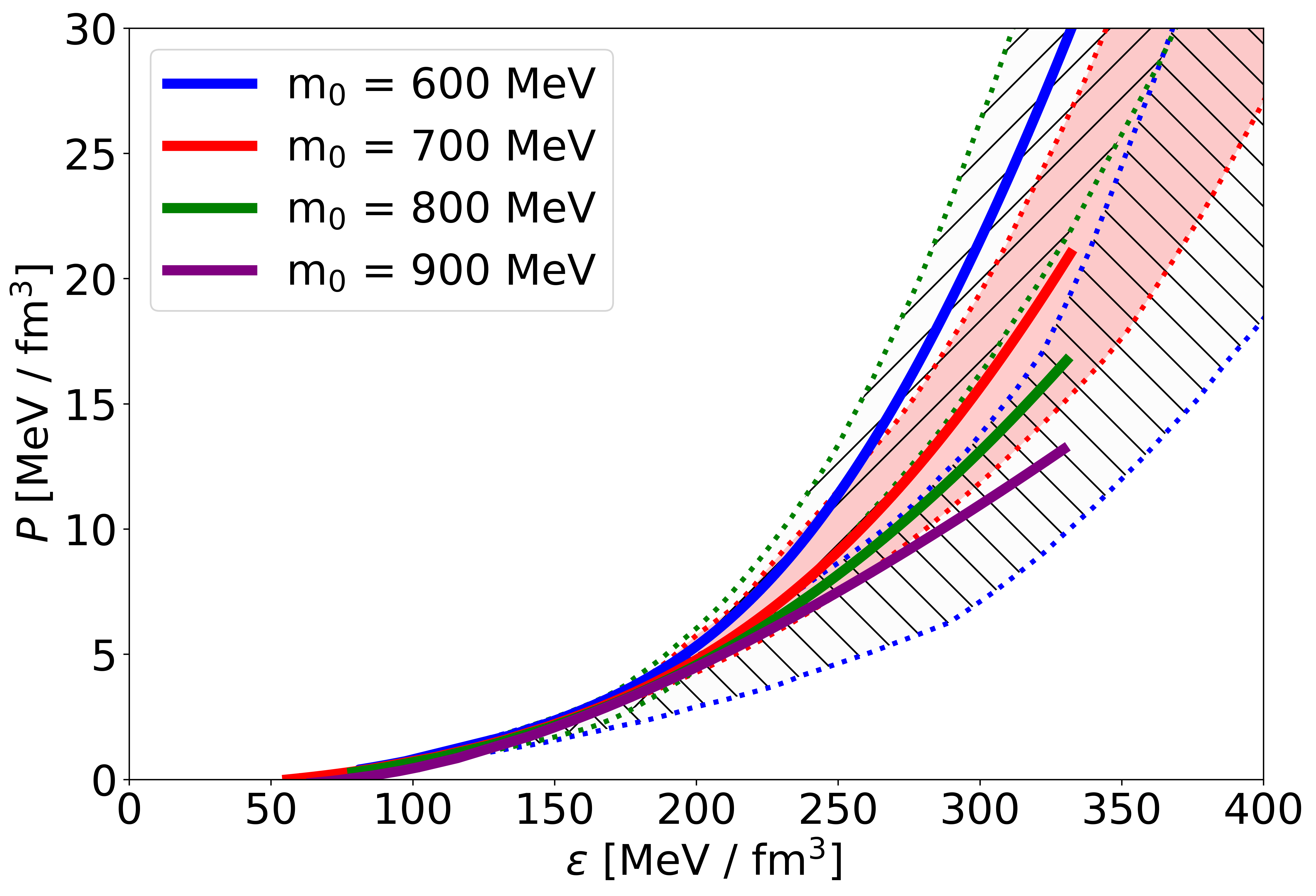}
\caption{Constructed EOS based on the parity doublet structure for several choices of chiral invariant mass $m_0=600, 700, 800, 900$ MeV with fixed slope parameter $L_0 = 57.7$ MeV. Colored band show  data-driven EOSs inferred from deep learning (edged by dotted red curves \cite{Fujimoto:2021zas}) and Bayesian (edged by blue dotted \cite{Raaijmakers:2021uju} and green dotted curves \cite{Ozel:2015fia,Bogdanov:2016nle}) analyses.}
\label{fig:EOS_compare}
\end{figure}

\textit{Construction of EOS.---}For the low density hadronic region ($ n_B \leq 2n_0$), we construct the EOS based on the parity doublet framework following \cite{Minamikawa:2020jfj,Minamikawa:2023eky}, where we have set the matching density at $n_L = 2n_0$. Below this density threshold, the hadronic description remains applicable, as other effects such as hyperons and many-body interactions do not significantly impact the EOS \cite{Bednarek:2011gd,Hebeler:2013nza}. The parameters inside the model are determined by fitting to physical inputs in vacuum (Pion decay constant $f_\pi=93$ MeV and hadron masses) and normal nuclear matter properties (Saturation density $n_0 \approx 0.16$ fm$^{-3}$ , binding energy $B_0 \approx 16$ MeV, incompressibility $K_0 \approx 240$ MeV, and symmetry energy $S_0 
\approx 31$ MeV). The model also incorporates vector meson $\omega$-$\rho$ mixing interaction, which significantly impacts the slope parameter $L_0$ of the symmetry energy. We systematically explore the parameter space by varying the chiral invariant mass $m_0$ with the slope parameter $L_0$ fixed to be 57.7 MeV consistent with \cite{Li:2021thg}. For each choice of chiral invariant mass $m_0$, we obtain a distinct EOS with different stiffness characteristics: larger values of $m_0$ lead to softer EOSs -- smaller pressure at the same energy density -- in the hadronic phase as shown in Fig.~\ref{fig:EOS_compare}.
The resulting EOS provides the low-density anchor for our unified NS EOS, which will be connected to higher densities as described in subsequent sections.
 
The quark matter EOS at high densities ($n_B \geq 5n_0$) is constructed using the NJL-type quark model \cite{Hatsuda:1994pi,Baym:2017whm, Baym:2019iky, Kojo:2021wax, Minamikawa:2020jfj,Yuan:2023dxl, Gao:2024chh, Gao:2024lzu,Gholami:2024ety}. In such density regime, quarks become the basic degrees of freedom, however, the strong coupling constant $\alpha_s \sim \mathcal{O}(1)$, making perturbative QCD calculation unrelaible. Our NJL-type effective model of quarks including the four-Fermi interactions which cause the spontaneous chiral symmetry breaking and the color-superconductivity. The EOS are controlled by two essential coupling constants: $g_V$ which controlls the repulsive vector interactions, and $H$ which governs attractive diquark correlations responsible for color superconducting phases. Following \cite{Baym:2017whm, Kojo:2021wax}, we treat these two couplings as phenomenological parameters which we will constrain later.

For the intermediate density region ($2n_0 \leq n_B \leq 5n_0$), we employ a model-independent approach based on fundamental physical principles following \cite{Komoltsev:2021jzg,Kurkela:2024xfh}. This method connects the low density EOS at the lower boundary ($P_L, n_L, \mu_L$ with $n_L = 2n_0$) to the NJL model at the upper boundary ($P_H, n_H, \mu_H$ with $n_H = 5n_0$) without making specific assumptions about the intermediate physics. Here $P_L$ and $\mu_L$ denote the pressure and baryon chemical potential from the low density EOS at $n_B = 2n_0$, while $P_H$ and $\mu_H$ correspond to those obtained from the NJL-type quark model at $n_B=5n_0$. The construction relies on two fundamental physical requirements: causality ($c_s^2 \leq 1$) and thermodynamic stability. Following the approach of \cite{Komoltsev:2021jzg}, these constraints, together with the requirement that the EOS must connect the low- and high-density boundary conditions, define a bounded region in the pressure-energy density ($P$-$\varepsilon$) plane. The upper boundary of this region represents the stiffest possible equation of state, while the lower boundary represents the softest, with both satisfying all physical requirements. This model-independent approach provides robust constraints on the intermediate density EOS without requiring detailed knowledge of microscopic physics in the transition region, naturally accommodating various quark-hadron transition scenarios without presupposing their specific nature.

\begin{figure}[htbp]\centering
\includegraphics[width=1\hsize]{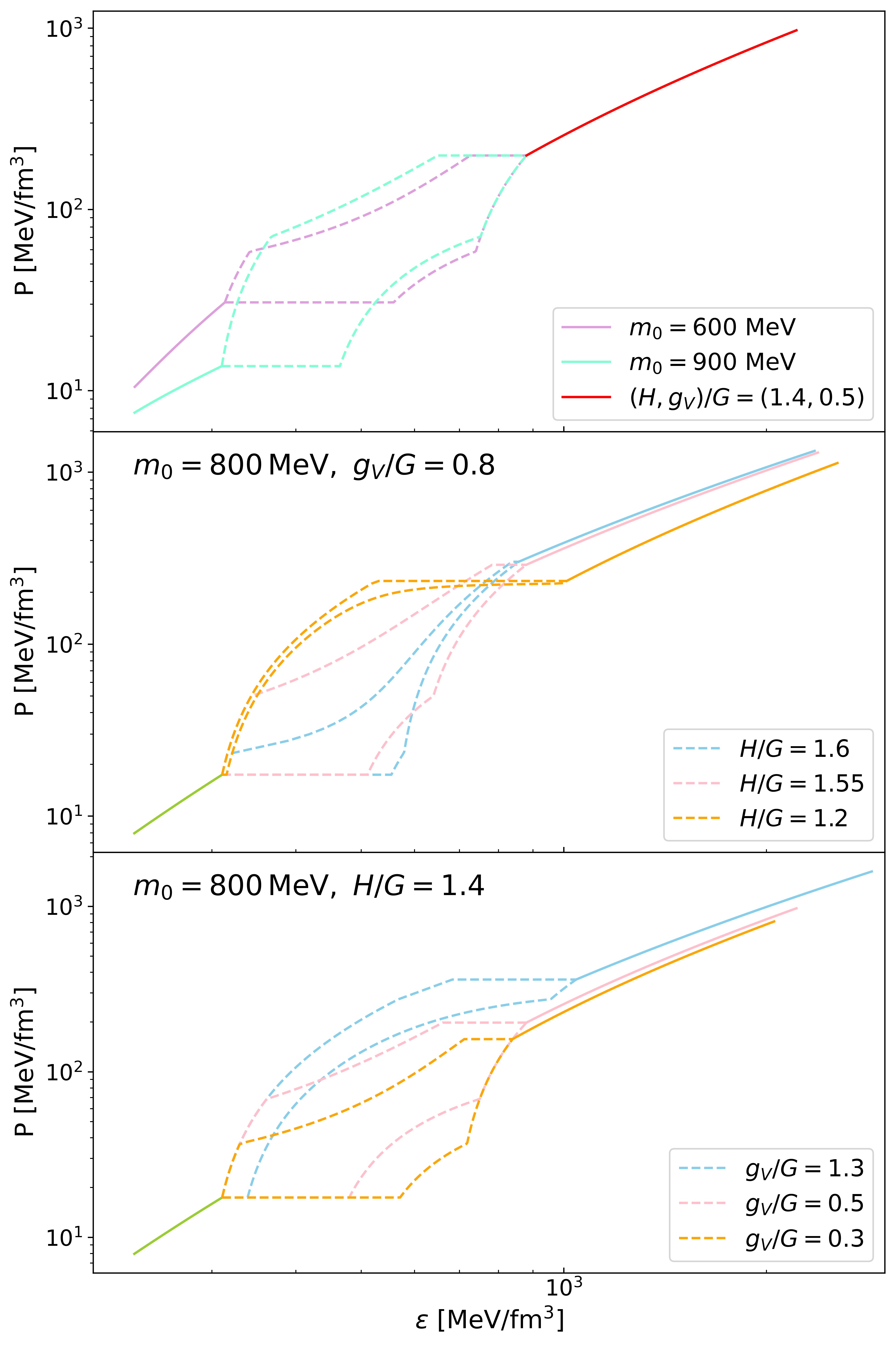}
\caption{Allowed regions in the pressure-energy density plane constrained by low-density hadronic EOS and high-density quark EOS with different choice of parameters. }
\label{fig:p_e_parameter}
\end{figure}

\begin{figure}[htbp]\centering
\includegraphics[width=1\hsize]{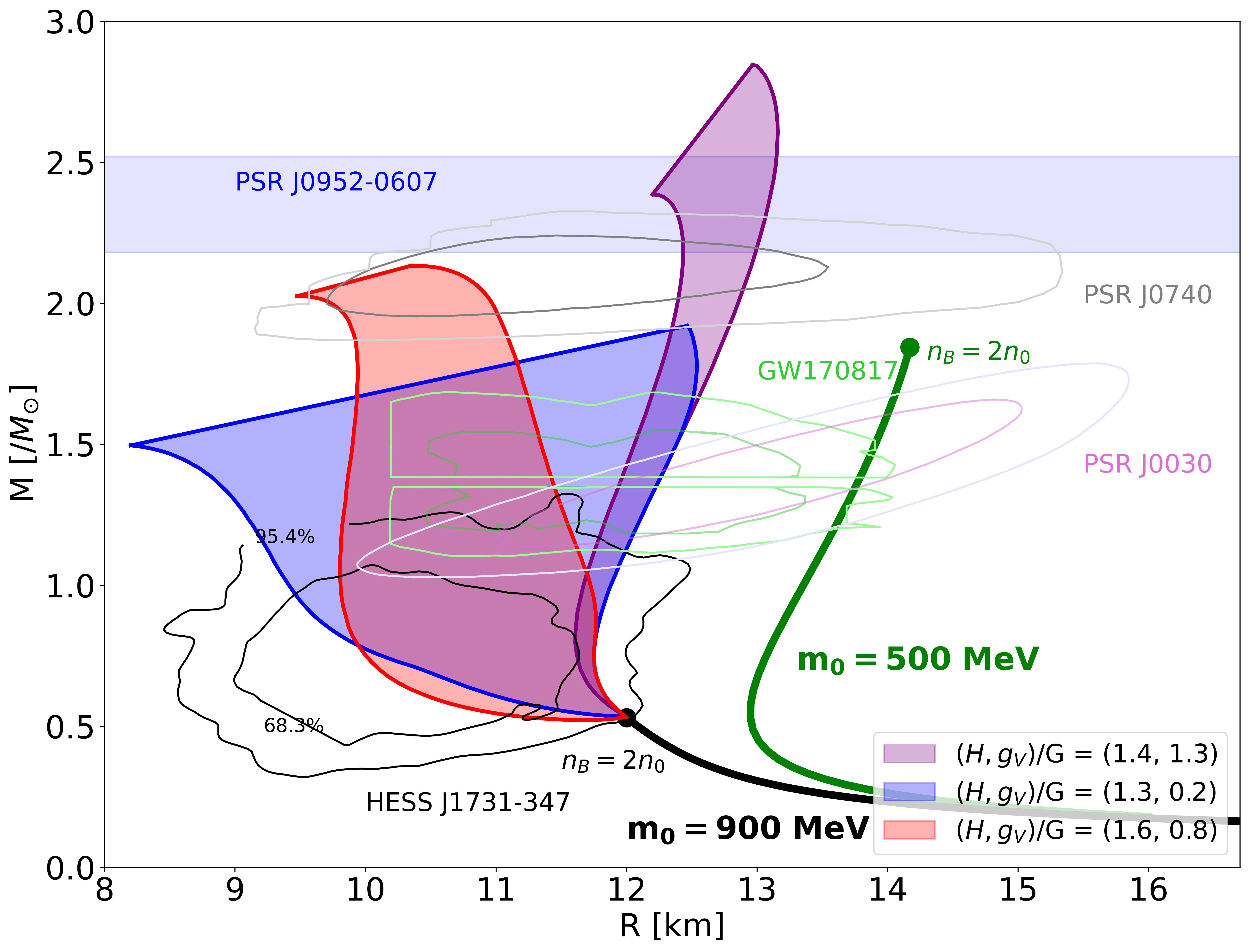}
\caption{Stable $M$-$R$ curves for NSs derived from our model with different $m_0$ values and $(H, g_V)/G$ combinations.Observational constraints are shown from: NICER measurements of PSR J0030+0451 \cite{Miller:2019cac} and PSR J0740+6620 \cite{Miller:2021qha}; PSR J0437-4715 \cite{Choudhury:2024xbk}; gravitational wave event GW170817 \cite{LIGOScientific:2018cki}; "black widow" pulsar PSR J0952-0607 \cite{Romani:2022jhd}; and the compact object HESS J1731-347 \cite{Doroshenko:2022nwp}.}
\label{fig:mr_final}
\end{figure}

\textit{Numerical results.---}
We systematically investigate how model parameters affect the NS EOS. Our framework involves two sets of key parameters: the chiral invariant mass $m_0$ for the the low-density hadronic phase, and the ($H, g_V$) couplings in the NJL-type quark model for the high-density quark phase.

Fig.~\ref{fig:p_e_parameter} examines the parameter dependencies in the pressure-energy density plane. For each parameter set, the dashed curves define boundaries of thermodynamically consistent regions connecting hadronic matter and quark matter, with upper and lower boundaries representing the stiffest and softest possible connections, respectively. In Fig.~\ref{fig:p_e_parameter}, the top panel shows the effect of varying $m_0$ with fixed quark matter parameters $(H, g_V)/G = (1.4, 0.5)$. Larger $m_0$ values lead to softer hadronic EOSs at low densities. Since all curves must connect to the identical NJL EOS at $n_B = 5n_0$, the boundaries for $m_0=900$ MeV become stiffer in part of the intermediate density region compared to the $m_0=600$ MeV case in order to maintain thermodynamic consistency. The middle panel demonstrates the effect of varying diquark coupling $H$ with fixed vector coupling $g_V/G = 0.8$. Increasing $H$ stiffens the quark matter EOS at high densities but softens the constrained boundaries at intermediate densities (below $\sim$700 MeV/fm$^3$), reflecting thermodynamic requirements when connecting to the hadronic EOS. The lower panel shows the effect of varying vector coupling $g_V$ with fixed diquark coupling $H/G = 1.4$. Larger $g_V$ values consistently stiffen the EOS across all densities, both in the pure quark matter region and throughout the allowed boundaries. This systematic stiffening directly impacts NS properties, resulting in both larger radii and higher maximum masses.

\textit{NS Properties and Parameter Constraints.---}
Having analyzed the effects of these parameters, we now examine their implications for NS structure. Each combination of $m_0$ and NJL parameters defines unique upper and lower boundaries in the pressure-energy density plane. These boundaries are constrained by causality and thermodynamic consistency requirements.

\begin{figure*}[htb]
\centering
\includegraphics[width=0.48\textwidth]{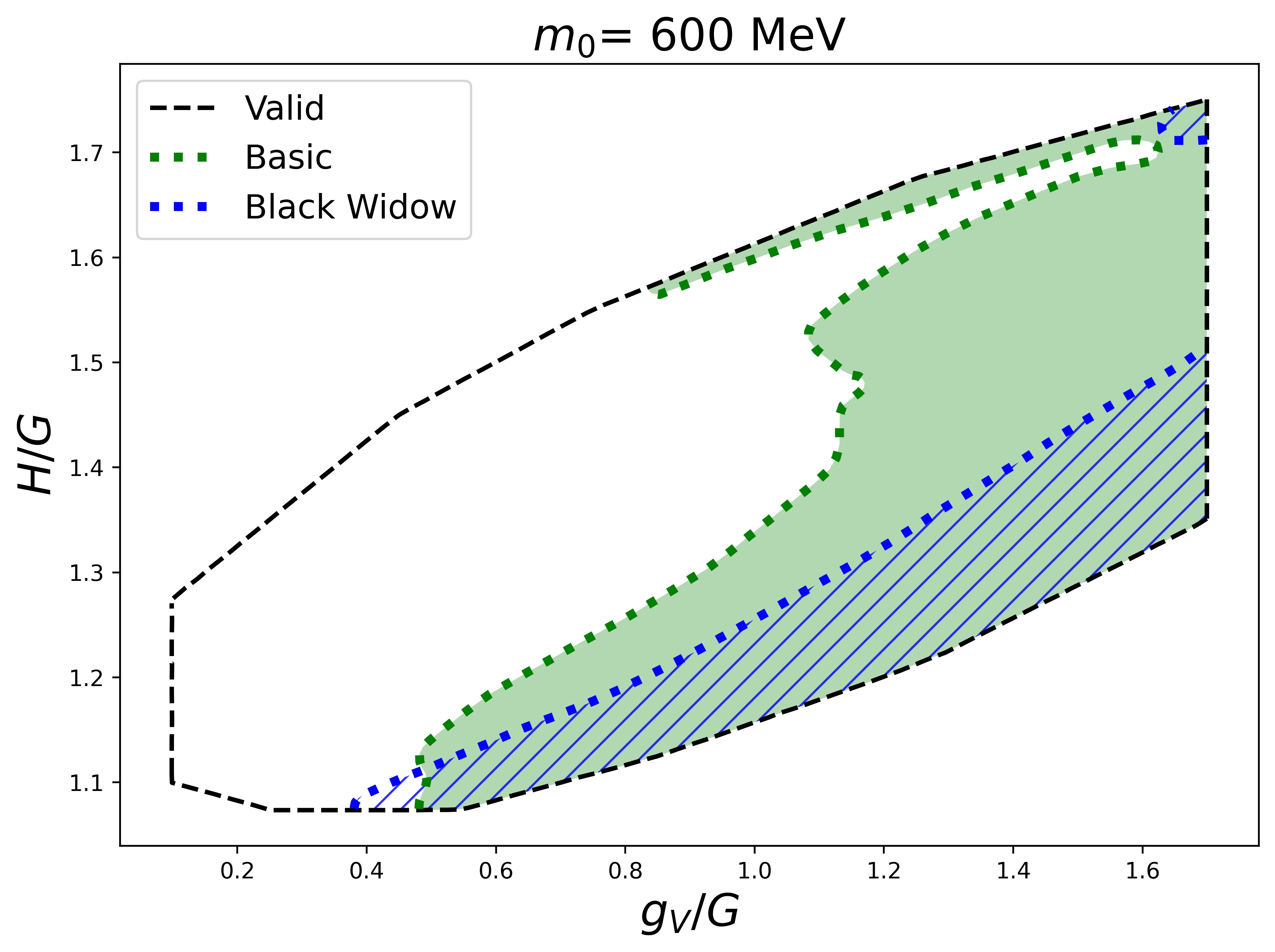}
\includegraphics[width=0.48\textwidth]{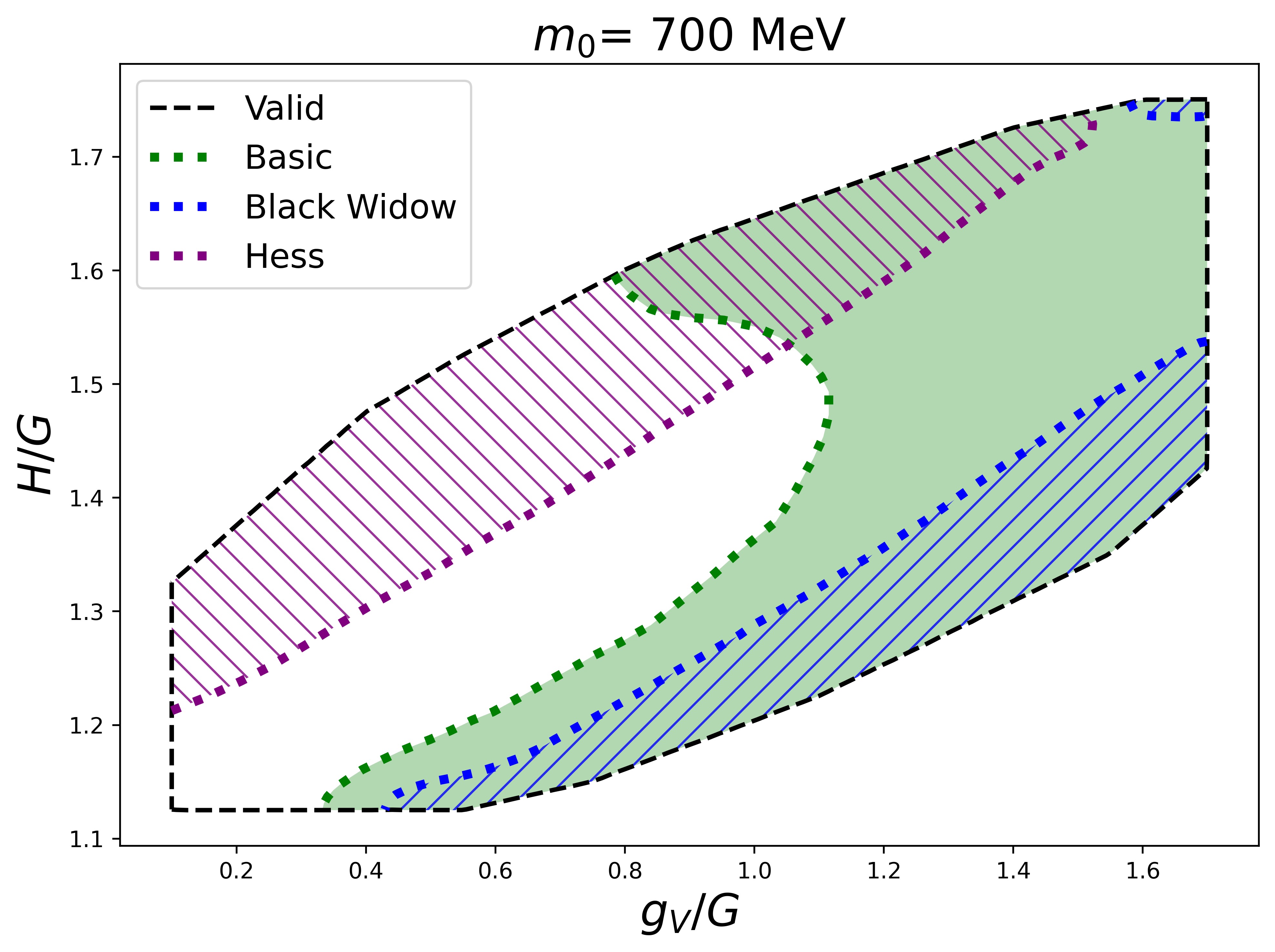}\\
\includegraphics[width=0.48\textwidth]{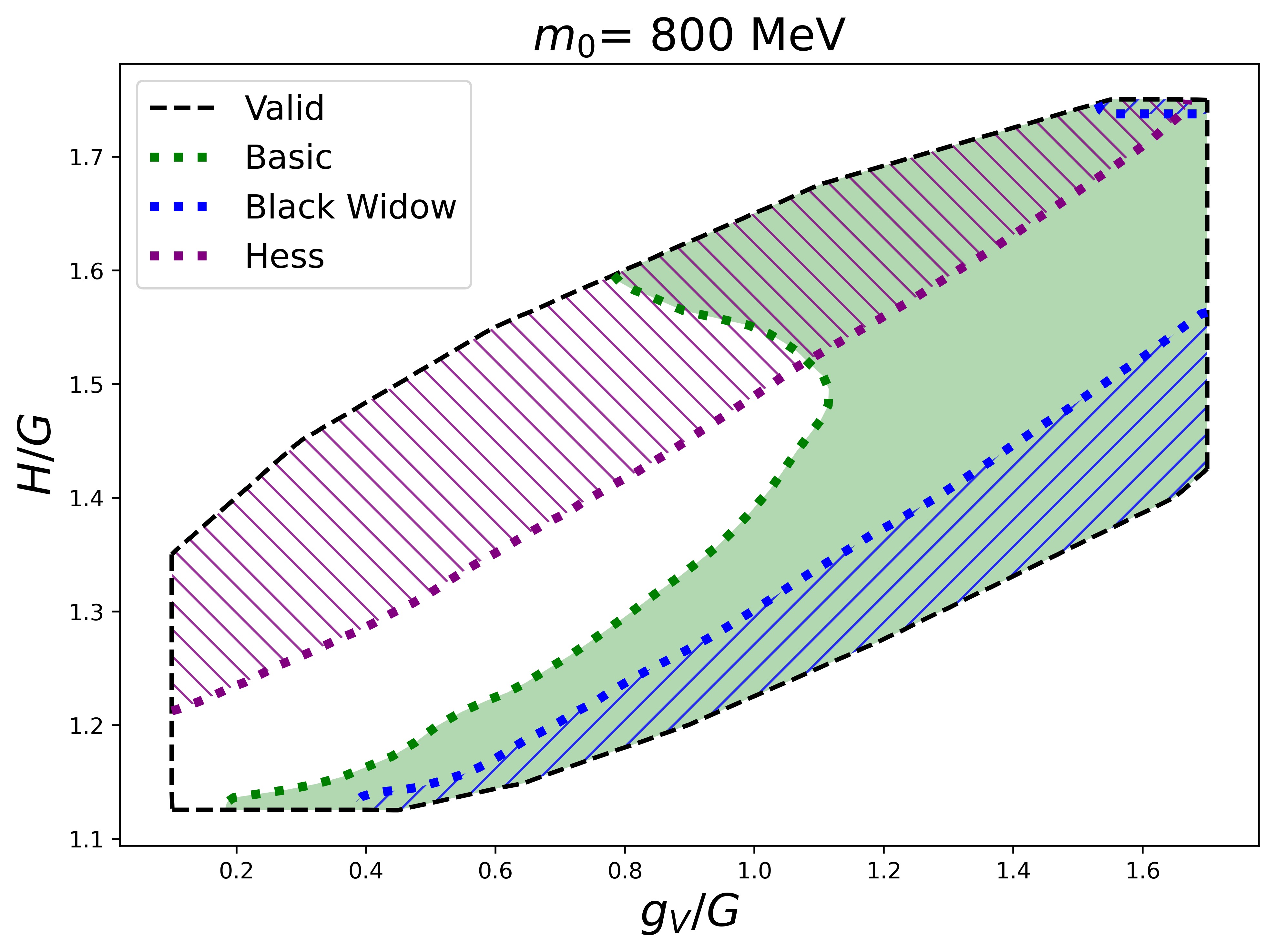}
\includegraphics[width=0.48\textwidth]{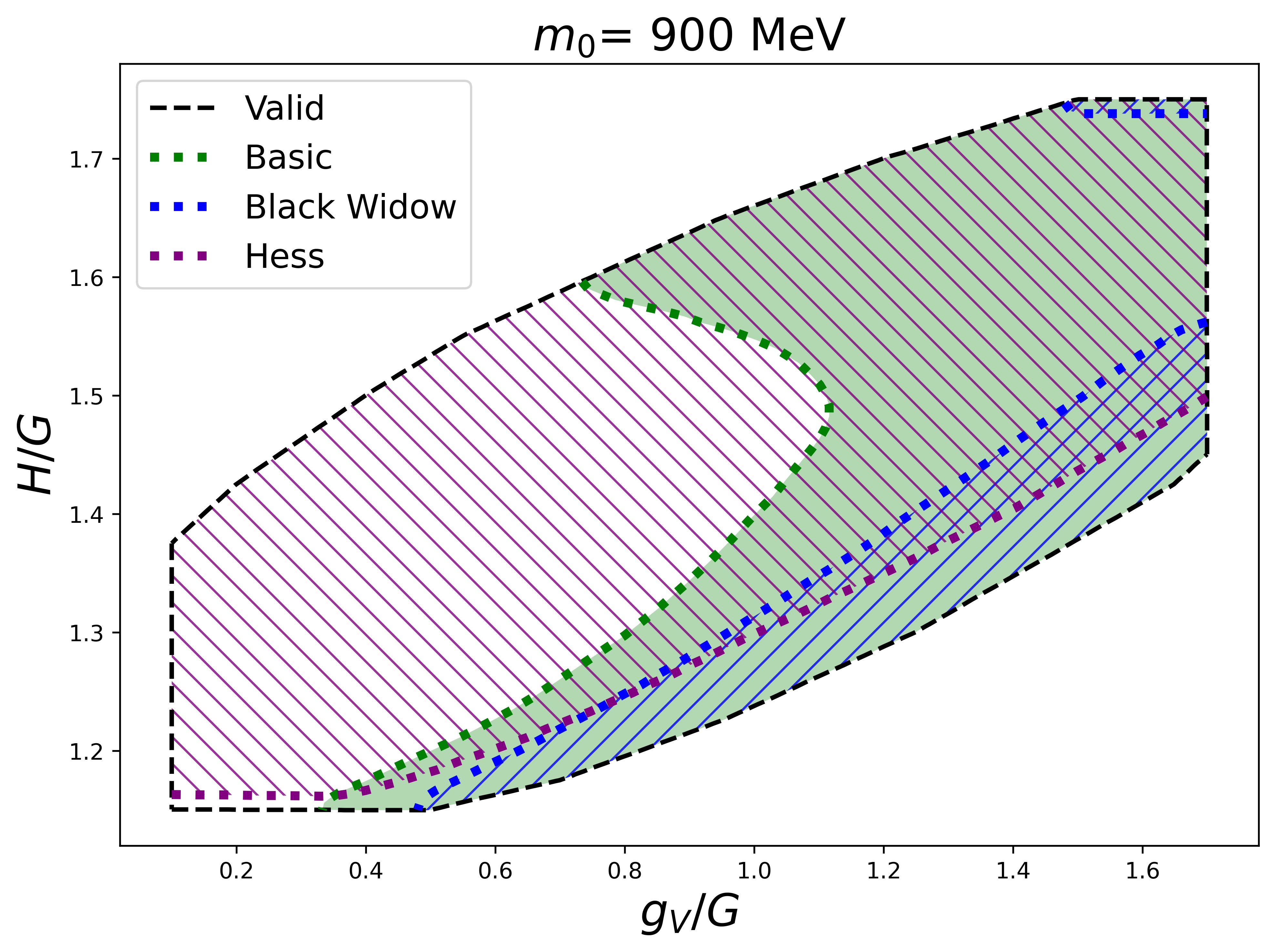}
\caption{
Constraints on the NJL-type quark model parameters $H$ and $g_V$ for four values of the chiral invariant mass $m_0$: 
(a) 600 MeV, (b) 700 MeV, (c) 800 MeV, and (d) 900 MeV. 
Black dashed curves indicate thermodynamically consistent regions where hadronic and quark EOS can be connected. 
Colored and hatched regions show parameter sets compatible with observational constraints (see text for details).
}
\label{fig:parameter}
\end{figure*}

By solving the Tolman-Oppenheimer-Volkoff (TOV) equations \cite{Tolman:1939jz,Oppenheimer:1939ne} with these boundary EOSs, we obtain the $M$-$R$ relationships shown in Fig.~\ref{fig:mr_final}. We divide the NS observations into different sets. The basic observational set includes NICER measurements of PSR J0030+0451 \cite{Miller:2019cac} ($M=1.44^{+0.07}_{-0.07} \,M_{\odot}$, $R = 13.7^{+2.6}_{-1.5}$ km) and PSR J0740+6620 \cite{Miller:2021qha} ($M=2.08^{+0.07}_{-0.07} \,M_{\odot}$, $R = 13.7^{+2.6}_{-1.5}$ km), constraints from PSR J0437-4715 \cite{Choudhury:2024xbk}($M=1.418^{+0.037}_{-0.037} \,M_{\odot}$, $R = 13.6^{+0.95}_{-0.63}$ km), and constraint to the radius obtained from the
LIGO-Virgo\cite{LIGOScientific:2017vwq,LIGOScientific:2017ync,LIGOScientific:2018cki}. The additional observational set includes the massive "black widow" pulsar PSR J0952-0607 \cite{Romani:2022jhd}($M=2.35^{+0.17}_{-0.17} \,M_{\odot}$) and the unusually light compact object HESS J1731-347 \cite{Doroshenko:2022nwp} ($M=0.77^{+0.2}_{-0.17} \,M_{\odot}$, $R = 10.04^{+0.86}_{-0.78}$ km). These  observations provide complementary constraints on our parameter space, allowing us to narrow the possible range of chiral invariant mass values.

As we increase the chiral invariant mass $m_0$ from 500~MeV (green curve) to 900~MeV (black curve), the EOS becomes progressively softer, resulting in smaller NS radii for the same mass, as expected. For the case of $m_0 = 500$~MeV, the pure hadronic EOS alone can already support a $1.4~M_{\odot}$ NS, but the corresponding radius falls outside the $1\sigma$ constraints from GW170817. This indicates that values of $m_0$ smaller than 500~MeV are excluded, even without considering the transition to quark matter.

For larger values of $m_0$, we examine how different quark matter parameters affect the allowed regions. Taking $m_0 = 900$~MeV as an example (black curve), we connect this hadronic EOS with various quark EOSs at $n_B = 5n_0$. The results show distinct constraints depending on the quark model parameters. For the soft quark matter EOS with $(H, g_V)/G = (1.3, 0.2)$,  the blue constrained region cannot support NSs with masses above $2~M_{\odot}$. For $(H, g_V)/G = (1.6, 0.8)$ (red region), we can accommodate most observations including GW170817, PSR J0030+0451, PSR J0740+6620, and HESS J1731-347, but still remains too soft to explain the massive ``black widow'' pulsar PSR J0952-0607 with $M \sim 2.3~M_{\odot}$. Finally, for stiff quark matter EOS with $(H, g_V)/G = (1.4, 1.3)$, denoted by the purple region, this combination can satisfy all observational constraints simultaneously.

Using these observational constraints, we systematically analyze the NJL model parameter space with the determined hadronic EOS. Figure~\ref{fig:parameter} presents our analysis of the parameter space in the $(H,g_V)$ plane for different values of the chiral invariant mass $m_0 = 600, 700, 800$, and 900~MeV. For each choice of $m_0$, the black dashed curves define the boundary of the thermodynamically allowed region where the hadronic EOS can consistently connect with the quark matter EOS for the given $(H, g_V)/G$ parameter combinations. The green region indicates parameter combinations that, after connecting with the corresponding quark matter EOS, can satisfy the basic observational set. The blue region with skewed shaded region represents $(H, g_V)/G$ parameter combinations that can accommodate the ``black widow'' pulsar observation, while the purple region with skewed shaded region indicates parameter combinations that can satisfy the HESS J1731-347 observation constraints.

For $m_0 = 600$~MeV, we cannot find any $(H, g_V)/G$ combination that satisfies the HESS J1731-347 observation. As we increase the value of $m_0$, the allowed parameter combinations that can accommodate the HESS object become larger. However, for $m_0 = 700$~MeV, there is no overlapping region that can simultaneously satisfy the basic observational set, HESS J1731-347 observation, and the ``black widow'' observation. If the HESS J1731-347 compact object is later confirmed to be a NS, we can further exclude chiral invariant mass values smaller than 700~MeV. When $m_0$ is increased to 800~MeV, we find some $(H, g_V)/G$ regions around $(1.72, 1.6)$ that can satisfy all observations simultaneously. For $m_0 = 900$~MeV, there are even larger regions that satisfy all observational constraints.

\textit{Summary and Discussions.---}
In this work, we established constraints on the nucleon chiral invariant mass through a systematic analysis of NS observations. Our novel approach combined three complementary models: the low-density hadronic matter based on parity doublet structure, the NJL-type quark model for the high-density quark matter, and a model-independent analysis of the intermediate density region based solely on fundamental physical principles.

By systematically exploring the parameter space, we examined how the chiral invariant mass $m_0$ affects NS properties. Our analysis incorporated observational constraints from multiple sources, including NICER measurements of PSR J0030+0451 and PSR J0740+6620, measurements from PSR J0437-4715, tidal deformability data from GW170817, the massive "black widow" pulsar PSR J0952-0607, and the unusually light compact object HESS J1731-347.

Our results reveal to satisfy the basic observational set, $m_0$ needs to be larger than 600~MeV, and this requirement remains unchanged even when including the "black widow" observation. To accommodate both the basic set and the HESS J1731-347 object, $m_0$ must be larger than 700~MeV. Finally, if we require all observations to be satisfied simultaneously, the hadronic EOS must be sufficiently soft, requiring $m_0$ larger than 800~MeV. These constraints revealed that approximately 60-95\% of nucleon mass likely originates from sources beyond the chiral condensate. This finding challenges the conventional understanding that nucleon mass arises primarily from spontaneous chiral symmetry breaking and points toward a more complex origin potentially involving explicit gluon contributions~\cite{Hatta:2018sqd} or other mechanisms. Particularly significant is our finding that only when the chiral invariant mass approaches 800 to 900~MeV can our model simultaneously accommodate both very massive NSs and unusually light compact objects like HESS J1731-347. A key strength of our approach is that it achieves these constraints without making specific assumptions about the nature of the quark-hadron transition. We do not presuppose whether this transition is a sharp first-order phase transition, a smooth crossover, or some exotic scenario. This model-independent treatment represents a major strength of our analysis, providing robust constraints that arise solely from fundamental physical principles and observational data.

Finally, we note that our choice of density boundaries—hadronic EOS up to $2n_0$ and quark matter EOS from $5n_0$—is based on empirical considerations where hadronic descriptions remain reliable up to $2n_0$, while quark degrees of freedom should become relevant at $5n_0$. The specific choice of these boundaries affects the derived constraints on $m_0$. When the hadronic matter EOS range is reduced to $1.5n_0$, the constraints become less restrictive: to satisfy all observational data (including HESS J1731-347 and the "black widow" pulsar), $m_0$ is constrained to be larger than 700 MeV. Similarly, raising the starting density of quark matter relaxes the constraint.

\begin{acknowledgments}
The authors gratefully acknowledge helpful conversations with Toru Kojo. This work is supported in part by JSPS KAKENHI Grant Nos.~ 23H05439, 24K07045 and JST SPRING, Grant No. JPMJSP2125. Y. L. M. is supported in part by the National Science Foundation of China (NSFC) under Grant No. 12347103, the National Key R\&D Program of China under Grant No. 2021YFC2202900 and Gusu Talent In novation Program under Grant No. ZXL2024363.
\end{acknowledgments}

\bibliography{sample631}

\end{document}